# Observation of the quantum Hall effect in confined films of the three-dimensional Dirac semimetal $Cd_3As_2$

Timo Schumann[1,†,*], Luca Galletti[1,†], David A. Kealhofer[2], Honggyu Kim[1], Manik Goyal[1], and Susanne Stemmer[1,*]

[1]Materials Department, University of California, Santa Barbara, CA 93106-5050, USA.

[2]Department of Physics, University of California, Santa Barbara, CA 93106-9530, USA.

[†]These authors contributed equally to this work.

[*]Email: schumann.timo@gmx.net; stemmer@mrl.ucsb.edu

**Abstract**

The magnetotransport properties of epitaxial films of $Cd_3As_2$, a paradigm three-dimensional Dirac semimetal, are investigated. We show that an energy gap opens in the bulk electronic states of sufficiently thin films and, at low temperatures, carriers residing in surface states dominate the electrical transport. The carriers in these states are sufficiently mobile to give rise to a quantized Hall effect. The sharp quantization demonstrates surface transport that is virtually free of parasitic bulk conduction and paves the way for novel quantum transport studies in this class of topological materials. Our results also demonstrate that heterostructuring approaches can be used to study and engineer quantum states in topological semimetals.

The realization of the quantum Hall effect, a hallmark of high-mobility, two-dimensional electronic systems, in the recently discovered three-dimensional Dirac and Weyl semimetals is of considerable scientific interest. These materials have unique surface states that reflect the topological properties of the crossings of the bulk conduction and valence bands at the Dirac or Weyl points [1]. The quantum Hall effect has recently been observed in three-dimensional topological insulators [2-5]. While quantum oscillations involving surface states have been reported for a three-dimensional Dirac semimetal [6], their connection to the bulk states prohibits the observation of the quantum Hall effect.

$Cd_3As_2$ is a three-dimensional topological Dirac semimetal with a single pair of Dirac cones that are displaced from the zone center and protected by the four-fold rotation symmetry in the tetragonal unit cell ($I4_1/acd$ space group) [7-13]. $Cd_3As_2$ hosts topologically non-trivial surface states [7, 12, 14]. To observe the quantum Hall effect from surface states, parallel conduction by the bulk carriers must be eliminated. This can be achieved, for example, by symmetry breaking and confinement in thin films and heterostructures, which are predicted to open energy gaps in the bulk states at the Dirac points [7, 15-18]. The generic surface states in Dirac semimetals, such as $Cd_3As_2$, are two sets of Fermi arcs, which are connected near the Dirac nodes through the bulk (see Fig. 1). These states are predicted to be robust, even if a gap opens in the bulk, and can give rise to quantum oscillations [19, 20].

In this Letter, we report on the magnetotransport properties of (112)-oriented, epitaxial $Cd_3As_2$ films grown by molecular beam epitaxy (MBE). We show that an energy gap opens in the bulk electronic states and, at low temperatures, carriers residing in surface states are sufficiently mobile to give rise to a quantized Hall effect.

Cd$_3$As$_2$ films were grown by molecular beam epitaxy on (111)B oriented GaAs substrates with 140-nm-thick (111)GaSb buffer layers. This approach yields high mobility, epitaxial films with the tetragonal (112) planes parallel to the (111) substrate/buffer layer surfaces [21]. Details of the growth procedure and structural characterization can be found in ref. [21]. The GaSb buffer layer is insulating and does not contribute to the transport [21]. The Cd$_3$As$_2$ films studied here were thinner than those in our prior investigations [21]. A cross-section transmission electron microscopy image of the heterostructure is shown in the Supplementary Information (Fig. S1 [22]). Hall bar structures, with widths of 50 μm and lengths of 100 μm were defined by standard photolithography and argon ion dry etching. Au/Pt/Ti layers served as ohmic contacts. Electrical measurements between 2 K and 300 K were performed in a Quantum Design Dynacool cryostat with magnetic fields up to 9 T and in an Oxford Instruments Triton dilution refrigerator at 1 K, in magnetic fields of up to 14 T. A small systematic offset in the values obtained from the electronics used for measuring resistances in the Oxford Instruments refrigerator was corrected using reference measurements performed in the Dynacool instrument.

The Hall mobility exceeds 1.3 m$^2$/Vs at low temperatures and the carrier density is 2.8×10$^{11}$ cm$^{-2}$. Compared to single crystals of Cd$_3$As$_2$ studied in the literature [6, 8, 13, 23] the corresponding three-dimensional carrier density is an order of magnitude (or more) lower. Thicker films grown under similar conditions show evidence for the presence of both p- and n-type carriers, which indicates that their Fermi level is close to the Dirac node (Fig. S2 in the Supplementary Information [22]).

Figure 2 shows the longitudinal ($R_{xx}$) and transverse ($R_{xy}$) resistances measured at 1 K as a function of magnetic field $B$ using a Hall bar structure. $R_{xx}$ features well-resolved Shubnikov-de Haas oscillations. Minima in $R_{xx}$ coincide with quantum Hall plateaus in $R_{xy}$. Their sequence is given by $R_{xy} = h/\nu e^2$, where ν is the filling factor, i.e., the number of conducting edge channels in the Landau-Büttiker picture of the quantum Hall effect [24]. Well-developed plateaus for $\nu = 2$ and $\nu = 1$ are visible and the measured resistances closely match the expected values of ~12.6 kΩ and 25.8 kΩ. $R_{xx}$ approaches zero at the plateaus, with only a small amount of residual resistance. Hall plateaus corresponding to higher filling factors are less well resolved and are presented in Fig. 3a. They are easier to detect in the derivative of the Hall resistance (Fig. 3b). While plateaus corresponding to $\nu = 4$ and $\nu = 6$ appear, there are none for $\nu = 3$ and $\nu = 5$.

Figure 3c shows the maxima in $R_{xx}$, which correspond to the Landau level crossing the Fermi energy and hence the transition between the filling factors ν, as a function of temperature. Oscillations are visible even at 20 K. A shoulder in the peak at ~ 4 T is clearly visible at low temperatures. The most straightforward assignment of the Landau levels is as follows. The Landau levels are two-fold degenerate at low B, which leads to the transition $\nu = 6 \rightarrow \nu = 4$ and the absence of $\nu = 5$. At higher B, lifting of the degeneracy leads to a shoulder for the $n = 1$ Landau level, but since $n_+ = n_- = 1$ still overlap, $\nu = 3$ is not resolved. The degeneracy is completely lifted for the lowest Landau level, resulting in the transition from $\nu = 2$ to $\nu = 1$. This assignment is used to construct the fan diagram, shown in Fig. 3d (the shoulder of the second peak was assigned to $n = 1$). The positions of the peaks in the fan diagram follow a straight line, supporting these Landau level assignments. We will discuss the non-zero intercept and the Landau level degeneracy below.

The observation of a quantized Hall effect by itself provides conclusive evidence of the two-dimensional nature of the electronic states. The fact that it is observed in a *three-dimensional* Dirac semimetal is remarkable and requires explanation. As discussed next, the temperature dependence of the carrier densities extracted from low-field Hall measurements and from the fan diagram show that the bulk spectrum is gapped and that the observed transport is only through the conductive surface states.

As shown in Fig. 4, the temperature dependence of the carrier density extracted from low-$B$ Hall measurements can be described (dotted line) by a two carrier model, where one set is thermally activated across a band gap of width $E_g$, while the density of a second set of carriers is temperature-independent. The thermally activated carriers reside in the gapped ($E_g \approx 35$ meV) bulk, while the second set is attributed to the gapless surface states. At high temperatures, bulk carriers dominate transport. They freeze out at low temperatures and transport becomes dominated by the carriers in the gapless surface states. This picture is further confirmed by comparing the carrier density measured by the Hall effect with the two-dimensional carrier concentration determined from the quantum oscillations. As shown in Fig. 4, the latter is independent of temperature within the whole temperature range where quantum oscillations can be detected and comparable to that extracted from the Hall measurements. Therefore, at low temperatures, all carriers are in the two-dimensional surface channels, where they give rise to Shubnikov-de Haas oscillations and quantized Hall plateaus. The behavior can be contrasted with that of thicker $Cd_3As_2$ films, reported in ref. [25], where the carrier density is higher and the density becomes independent of temperature already at 100 K, consistent with most carriers residing in bulk Dirac semimetal states. Thicker films do not show the quantum Hall effect (see also Supplementary Information, Fig. S2 [22]).

Pronounced negative magnetoconductance ($\Delta G_{xx}$) is observed (Fig 5a) at low $B$, which can be fitted (dashed lines) over a wide temperature range (2 K – 75 K), using a model for weak anti-localization [26]:

$$\Delta G_{xx} = \alpha \frac{e^2}{h} \left( \Psi \left[ \frac{B_\varphi}{B} + \frac{1}{2} \right] - \ln \left[ \frac{B_\varphi}{B} \right] \right) + \beta B^2 \qquad (1)$$

where $\Psi$ is the digamma function, $L_\varphi = \sqrt{\frac{h}{8\pi e B_\varphi}}$ is the phase coherence length, $e$ is the elementary charge, and $h$ is Planck's constant. The $\beta B^2$ term accounts for a weak background in the conductivity and the factor $\alpha$ takes values between ½ (weak antilocalization) and -1 (weak localization). The fits show that $\Delta G_{xx}$ is well described by weak anti-localization. Extracted values for $\alpha$ and $L_\varphi$ are shown in Fig. 5b. The phase coherence length increases from 60 nm at 75 K to 180 nm at 2 K. The parameter α is positive for all temperatures and approaches 0.5 at low temperatures, which can be interpreted as arising from a Berry phase of π of a topological material [27, 28].

Another indicator of a Berry phase is a phase shift $\gamma$ in the Shubnikov-de Haas oscillations [29], which can be determined from the intercept in the fan diagram. Here, measurements at all temperatures yield an intercept of -0.28 (Fig. 3d), suggesting that the electronic states giving rise to the oscillations are topologically non-trivial. A Berry phase of π corresponds to $\gamma = 0$ and an intercept of ½ in the fan diagram for two-dimensional systems. Deviations from this value have been variously attributed. One explanation in $Cd_3As_2$, as well as other materials with large g-factors, is a departure from semi-classical models at low quantizing numbers $n$ [30, 31]. The specifics of the "Weyl" orbits (Fig. 1) should also play a role. Caution should therefore be applied in the interpretation of the Berry phase from the Landau plot.

The appearance of a bulk gap in conjunction with gapless surface states is consistent with predictions of topological state transitions due to the confinement of a three-dimensional Dirac semimetal [7, 15-17, 32, 33]. The observation of weak antilocalization and a non-zero intercept in the Landau fan diagram are consistent with a nonzero Berry phase, i.e. topological non-trivial surface states. Beyond this, however, the results in this study raise questions as to how to interpret the quantum Hall effect, which directly relates to their topological nature.

The Hall effect of massless Dirac fermions takes on a half-integer quantized form, i.e., $\nu = n + \frac{1}{2}$ [34, 35]. Half-integer quantum Hall plateaus cannot be observed, because the chiral edge states have to be integer-quantized [36, 37]. Odd integer sequences of $\nu$ in the quantum Hall effect in three-dimensional topological insulators have been taken as a signature of Dirac states of degenerate top and bottom surfaces [2, 3]. This is not what is observed here. A two-fold degenerate quantum Hall effect may be consistent with two copies "Weyl" orbits, as depicted in Fig. 1. The results indicate that the degeneracy of these orbits is lifted in high magnetic fields. Although further investigations are needed to explain this, it is likely that details of the surfaces are relevant, in analogy with valley splitting for semiconductor surfaces [38]. We note that any spin degeneracies should already be lifted at the high magnetic fields required to observe the quantum Hall effect, given the very large $g$-factor of $Cd_3As_2$ [23]. The Hall plateau sequences for the case of a quantum confined Dirac semimetal, as a function of microscopic parameters such as orientation and interface asymmetry, clearly need further theoretical investigation. Comparison with the experimental results presented here should shed light on the topological aspects of the surface states, such as degeneracy and expected value of the Berry phase.

In summary, we have shown that high-mobility thin films of a three-dimensional topological Dirac semimetal, $Cd_3As_2$, allow for engineering the topological states and associated intrinsic quantum transport phenomena. We showed that by reducing the dimensionality of a three-dimensional topological Dirac semimetal, $Cd_3As_2$, the bulk states are gapped out. This disconnects the surface and bulk states and a quantized Hall effect appears due to highly confined, gapless surface states with high mobility. The high carrier mobility and a clear quantized Hall achieved here are an important first step in the search for novel exotic states at high magnetic fields [39] and at interfaces [40] in this new class of topological semimetals.

**Acknowledgements:**

The authors thank Andrea Young, Leon Balents, Yuanming Lu, and Jim Allen for discussions and Yuntian Li for assistance with the measurements. They also gratefully acknowledge support through the Vannevar Bush Faculty Fellowship program by the U.S. Department of Defense (grant no. N00014-16-1-2814). Partial support was also provided by a grant from the U.S. Army Research Office (grant no. W911NF-16-1-0280). The dilution fridge used in the measurements was funded through the Major Research Instrumentation program of the U.S. National Science Foundation (award no. DMR 1531389). This work made use of the MRL Shared Experimental Facilities, which are supported by the MRSEC Program of the U.S. National Science Foundation under Award No. DMR 1720256.
9

**Figure Captions**

**Figure 1:** Generic picture of Fermi arcs on the surfaces (yellow) of a three-dimensional Dirac semimetal film with a bulk gap, according to theoretical predictions [19, 33]. The surface arcs connect to projections of the Dirac nodes, leading to two copies of "Weyl orbits" and a connection through the bulk, indicated by the dashed lines [19, 33]. Asymmetry in top/bottom surfaces and other microscopic parameters will determine specifics, such as Fermi arc shapes (as shown here, after ref. [33]).

**Figure 2:** Quantum Hall effect in a 20-nm-thick epitaxial $Cd_3As_2$ film grown by molecular beam epitaxy. Shown are the Hall ($R_{xy}$) and longitudinal ($R_{xx}$) resistances as a function of magnetic field measured at 1 K.

**Figure 3:** (a) Hall data from Fig. 2, plotted over a more limited $B$ field range, to show the presence/absence of higher filling factors $\nu$. (b) Derivative plot of the data shown in (a), which more clearly shows the emerging Hall plateaus for higher filling factors. Here, $\nu = 6$ and $\nu = 4$ are present, whereas $\nu = 3$ and $\nu = 5$ are absent. (c) Temperature dependence of the Shubnikov-de Haas oscillations with the Landau levels indexed as indicated. The data was recorded from the same sample, but a different device than that for the data shown in Fig. 2. (d) Fan diagram obtained by indexing the Landau levels as shown in (c). The dashed-dotted lines are linear fits.

**Figure 4:** Temperature dependence of the carrier densities and the Hall mobility. The carrier densities ($N$) were extracted from low field Hall coefficients [$N$ (Hall)] and from the Shubnikov-de Haas oscillations [$N$ (SdH)], respectively. The two measurements agree at low temperatures,

when the bulk carriers are frozen out and only the carriers in the surface states remain. The fit to a two carrier model (see text) is shown as a dotted line. Also shown is the Hall mobility ($\mu$).

**Figure 5:** (a) Conductance at low magnetic fields as a function of temperature. Fits to a weak antilocalization model are shown as dashed lines. (b) Extracted values for the phase coherence length and scaling parameter as a function of temperature. The data was measured on a sample with a thickness of 35 nm, which also showed the quantum Hall effect.

**Figure 1**

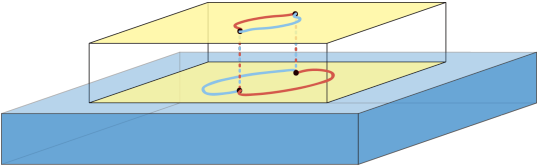

**Figure 2**

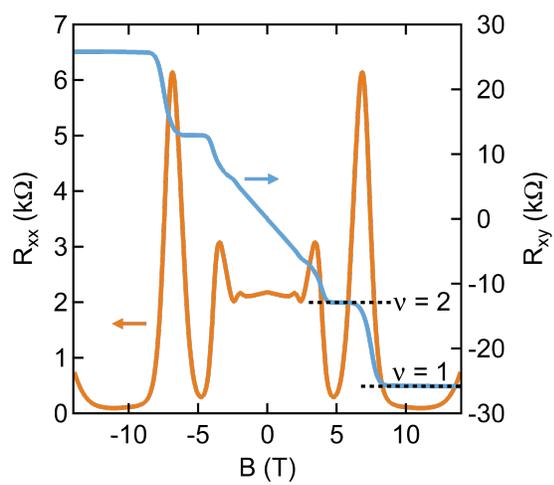

**Figure 3**

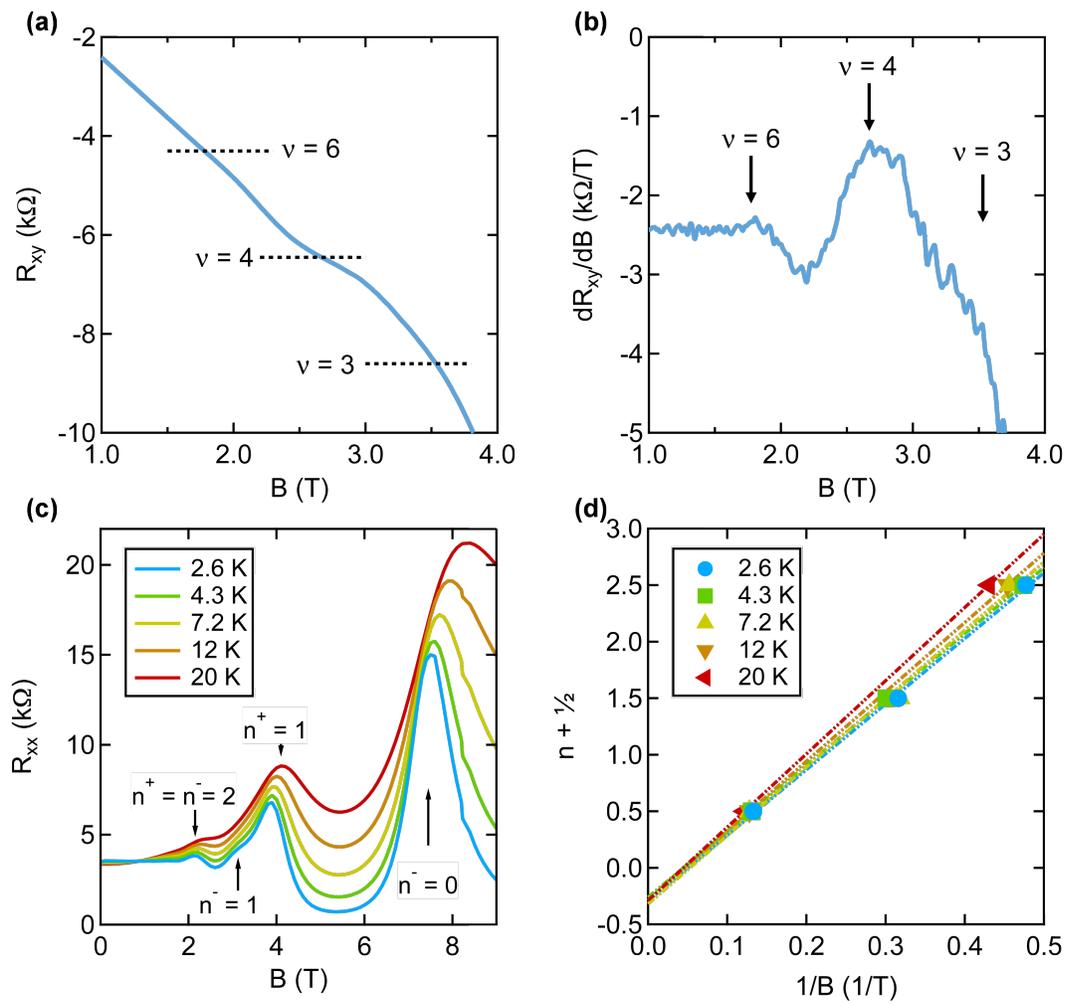

**Figure 4**

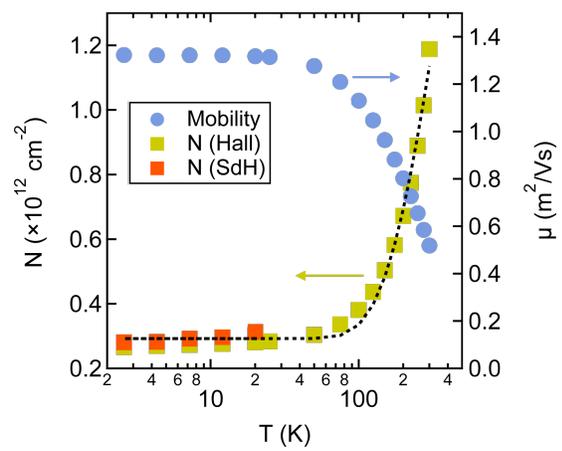

**Figure 5**

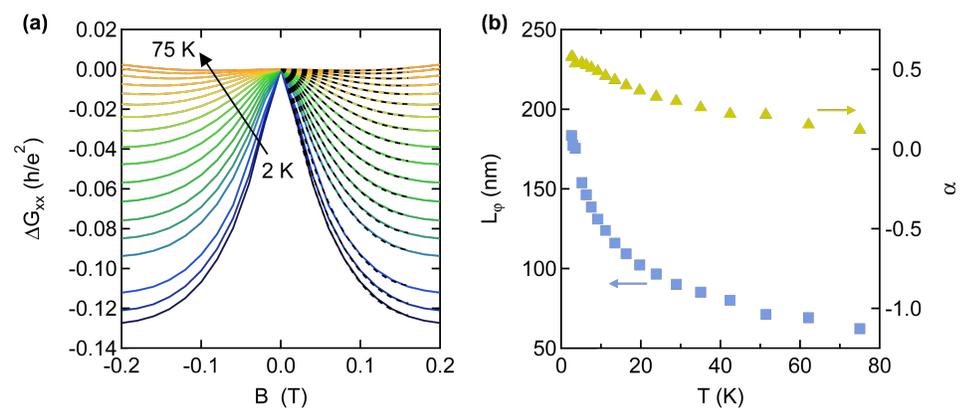